\documentclass{aa}                 %% \documentstyle[epsfig]{l-aa}
\usepackage{graphics}
\begin{document}

\thesaurus{9          % A&A Section 7: The Sun [=A&A Main Journ.]
           (06.18.1;  % Sun: radio radiation,
            06.06.3;  % Sun: flares,
            06.03.2;  % Sun: corona,
            02.16.1;  % Plasmas,
            02.13.2)} % Magnetohydrodynamics (MHD)

\title{Solar flare radio pulsations as a signature of dynamic magnetic
       reconnection}

\author{B.~Kliem \inst{1} \and M.~Karlick\'y \inst{2} \and A.O.~Benz 
        \inst{3}}

% \offprints{B.~Kliem}

\institute{Astrophysical Institute Potsdam, An der Sternwarte 16,
           D-14482~Potsdam, Germany; bkliem@aip.de
\and       Astronomical Institute, Academy of Sciences, 
           CS-251\,65~Ond\v{r}ejov, Czech Republic; karlicky@sunkl.asu.cas.cz
\and       Institute of Astronomy, ETH, CH-8092 Z\"urich, Switzerland;
           benz@astro.phys.ethz.ch}

\date{Received 28 September 1999 / Accepted 29 May 2000}
\authorrunning{B.~Kliem et al.}
\titlerunning{Decimetric solar flare radio pulsations}
\maketitle

\begin{abstract}

Decimetric radio observations of the impulsive solar flare on October 5,
1992, 09:25~UT show a long series of quasi-periodic pulsations deeply
modulating a continuum in the 0.6--2~GHz range that is slowly drifting
toward lower frequencies.  We propose a model in which the pulsations of
the radio flux are caused by quasi-periodic particle acceleration episodes
that result from a dynamic phase of magnetic reconnection in a large-scale
current sheet.  The reconnection is dominated by repeated formation and
subsequent coalescence of magnetic islands (known as ``secondary tearing''
or ``impulsive bursty'' regime of reconnection), while a continuously
growing plasmoid is fed by newly coalescing islands.  Such a model,
involving a current sheet and a growing plasmoid, is consistent with the
Yohkoh observations of the same flare (Ohyama \& Shibata \cite{OS98}).  We
present two-dimensional MHD simulations of dynamic magnetic reconnection
that support the model.  Within the framework of the proposed
interpretation, the radio observations reveal details of plasmoid
formation in flares. 

\end{abstract}

\section{Introduction}

Although the physical origins of a number of solar radio burst types are
well known, e.g., type II and type III bursts (Kr\"uger \cite{Kr79};
Bastian et al.\ \cite{Ba98}), there are still many questions which types of
the radio plasma emission emerge directly from the primary energy release
of solar flares, which is supposed to result from magnetic reconnection in
coronal current sheets (e.g., Priest \cite{Pr82}; Masuda et al.\
\cite{Ma94}; Tsuneta \cite{Ts96}).  Furthermore, there are still different
viewpoints regarding the question in which frequency range this radio
emission can be observed.

Decimetric continuum radio bursts are often related to the impulsive phase
of solar flares, and their frequencies correspond to the densities supposed
to exist in the primary energy release volume.  These bursts are usually
regarded to result from a distribution of flare-accelerated particles
trapped in a coronal magnetic loop (Kuijpers \cite{Ku80}).  In the trap,
the particles develop a loss-cone distribution function, which is unstable
and supports a plasma emission process that leads to fundamental emission
near the electron plasma frequency, $\omega_{pe}=(4\pi ne^2/m_e)^{1/2}$, or
to harmonic emission near $2\,\omega_{pe}$.  According to this model, the
bursts carry signatures not only of the particle acceleration in the
primary energy release process, but also of the evolution of the trapped
particle distribution in velocity space and of the evolution of the loop
during the flare.

Continuum radio bursts, especially those in the decimetric wavelength
range, often exhibit broad-band pulsations of the flux.  The timescales
lie typically in the range $\sim0.5\mbox{--}5$~s. Proposed mechanisms of
the pulsations fall into three categories:  (1) MHD oscillations of a
magnetic loop, which modulate the radio emissivity, (2) an oscillatory
nonlinear regime of the kinetic plasma instabilities that emit the radio
waves, and (3) a modulation of the acceleration of particles which
represent the source of the wave energy (Aschwanden \cite{Asch87}).  The
third of these includes non-stationary magnetic reconnection.

Improved knowledge of magnetic reconnection in current sheets is imperative
for resolving the flare problems.  Such studies can be guided by radio
observations, which often carry details of the dynamical plasma processes
not visible at other wavelengths.  In numerical simulations of reconnection
(e.g., Ugai \& Tsuda \cite{UT77}; Sato \& Hayashi \cite{SH79}; Karlick\'y
\cite{Ka88}; Scholer \cite{Scho91}; Ugai \cite{Ug92}), as well as in the
interpretation of soft X-ray observations (Tsuneta \cite{Ts96}), the focus
has usually been on stationary Petschek-type reconnection.  However,
temporal fine structure is abundant in radio bursts and often also present
in hard X-ray bursts (e.g., Kiplinger et al.\ \cite{Ki83}).  Another
problem lies in the fact that numerical simulations have to be performed in
rather small boxes and over rather short time intervals.  There is a
permanent need to extend the simulations further in space (where they tend
to reveal stronger dynamics) and in time, and attention must be devoted to
the re-scaling of the numerical results when specific observations are to
be interpreted.

In the present paper we consider a pulsating decimetric continuum radio
burst event.  Its peak was coincident with the impulsive phase of a solar
flare (October 5, 1992).  Rather detailed knowledge of this flare is
already available from a thorough analysis of its soft X-ray emission,
which includes, in particular, the ejection of a plasmoid (Ohyama \&
Shibata \cite{OS98}).  We argue that the pulsations can be explained by
modulations of particle acceleration in a highly dynamic regime of
magnetic reconnection in an extended current sheet above the soft X-ray
flare loop and that the radio emission process does not necessarily
require trapping in a magnetic loop.  The time profile of the radio burst
appears to reflect the dynamics of the reconnection process directly.  MHD
simulations of reconnection in a long current sheet are presented to
support the model. These simulations provide also further insight into the
process of plasmoid formation in a current sheet, which is directly
related to episodes of enhanced reconnection and enhanced particle
acceleration.

The observations are presented in Sect.~\ref{obs} and our MHD simulations
are presented and scaled to coronal conditions in Sect.~\ref{reconn}.  A
discussion of the new interpretation of pulsating continuum bursts and the
conclusions are given in Sects.~\ref{discuss} and \ref{concl},
respectively.

\section{Observations}
\label{obs}

\begin{figure}
 \begin{center}
  % \fbox{
   % \resizebox{0.84\textwidth}{!}{\includegraphics{fig/figu1.ps}}
   % \resizebox{\hsize}{!}{\includegraphics{fig/figu1.ps}}
   %%% \resizebox{\hsize}{!}{\includegraphics{fig/b921005.eps}}
   % \resizebox{\hsize}{!}{\includegraphics{fig/KK_main.ps}}
   \resizebox{\hsize}{!}{\includegraphics{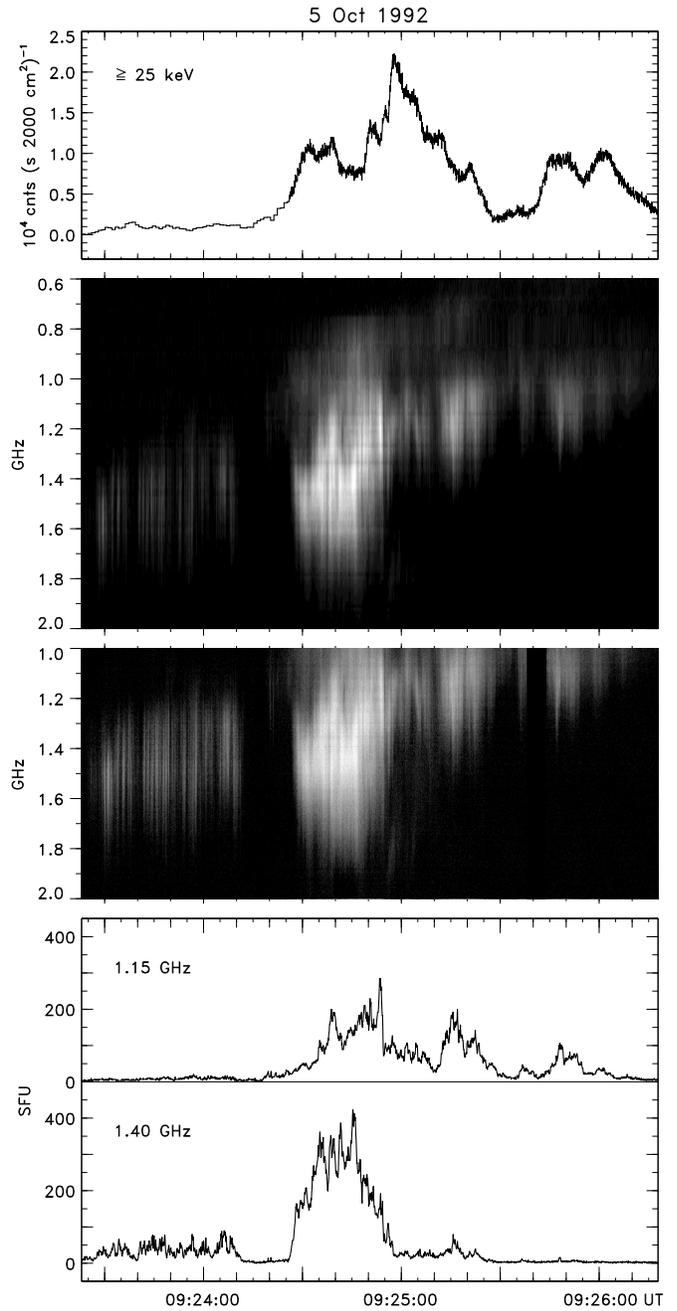}}
   % \parbox{11.5cm}{\hfill\vspace{10cm}}
  % }
  \caption{The hard X-ray lightcurve of the October~5, 1992 flare 
           (CGRO/BATSE/DISCSC data; top), 
           the dynamic radio spectrum (Z\"urich and Ond\v{r}ejov 
	   radiospectrographs; 2nd and 3rd panel, respectively), and 
	   two single-frequency cuts of the dynamic spectrum (bottom).
	   The Ond\v{r}ejov spectrum contains a gap of $\approx6$~s 
	   at 9:25:38~UT.}
  \label{main}
 \end{center}
\end{figure}

\begin{figure}
 \begin{center}
  % \fbox{
   % \resizebox{0.9\hsize}{!}{\includegraphics{fig/figu3.ps}}
   %%%\resizebox{0.9\hsize}{!}{\includegraphics{fig/a921005.eps}}
   % \resizebox{\hsize}{!}{\includegraphics{fig/KK_microwave.ps}}
   \resizebox{\hsize}{!}{\includegraphics{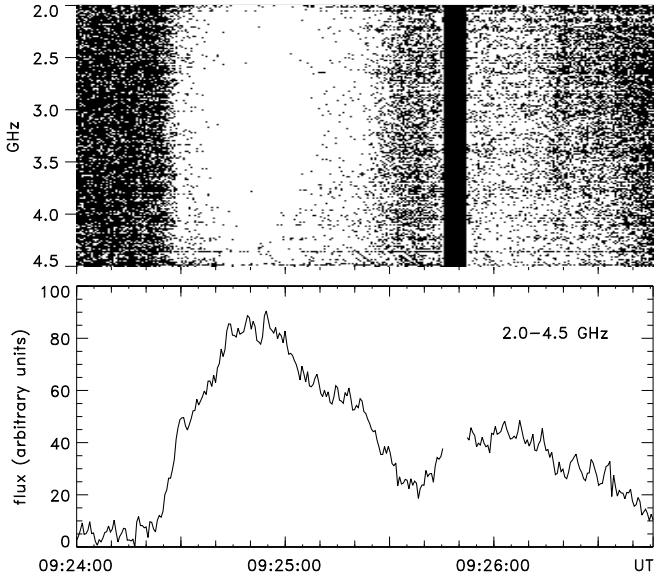}}
   % \parbox{8.5cm}{\hfill\vspace{5.5cm}}
  % }
  \caption{High frequency radio continuum coincident with the rise of the
           hard and soft X-ray emissions (Ond\v{r}ejov spectrograph).}
  \label{microwave}
 \end{center}
\end{figure}

The flare on October 5, 1992, 9:25~UT was an impulsive, compact-loop, M2 
class event that took place at the west limb, probably in active region
NOAA 7293, which was located slightly behind the limb. Its hard X-ray
lightcurve and the decimetric dynamic radio spectrum are plotted in
Fig.~\ref{main} along with two characteristic single-frequency
decimetric flux profiles. Fig.~\ref{microwave} shows the corresponding
microwave burst.

The hard X-ray (HXR) emission showed a complex temporal evolution. The
flare started with an extended period of low-level emission. The first
impulsive rise near 9:24:25~UT was followed by several strong
enhancements, each of $\ga10$~s duration (FWHM). The spectrum was hardest
near 9:25:05~UT. The Yohkoh/HXT 14--23\,keV and 23--33\,keV band images
(9:24:28--9:25:16~UT) show a single, slightly elongated source with
centroid position just above the west limb (Sato et al.\ \cite{Sa98}). The
weak microwave burst had a time profile similar to the HXR burst, but was
much smoother. Also the GOES soft X-ray lightcurves (Solar Geophysical
Data) show an impulsive rise near 9:24:25~UT, following a period of
low-level activity.

The decimetric radio burst consisted of continuum emission superimposed
with strong broadband pulsations during its whole lifetime. The general
flux level varied in a complex manner during the event, showing several
enhancements, as is typical of decimetric bursts. The burst as a whole
drifted towards lower frequencies at a rate $\approx-3$~MHz\,s$^{-1}$. The
main burst phase started at 9:24:27~UT at a frequency slightly higher than
the preceding pulsations and with a temporarily much higher drift rate of
$\approx-9$~MHz\,s$^{-1}$. The average drift rate decreased toward the end
of the event. The burst was weakly polarized, which is typical of type IV 
bursts from sources near the solar
limb. It is not known whether the emission was fundamental or harmonic
radiation. No other bursts were recorded at this time at frequencies below
0.6~GHz (Tremsdorf and Z\"urich radiospectrographs).

\begin{figure}[t]
 \begin{center}
  % \fbox{
   % \resizebox{0.84\textwidth}{!}{\includegraphics{fig/figu2.ps}}
   %%%\resizebox{\hsize}{!}{\includegraphics{fig/figu2.ps}}
   % \resizebox{\hsize}{!}{\includegraphics{fig/KK_pulsedetail.ps}}
   \resizebox{\hsize}{!}{\includegraphics{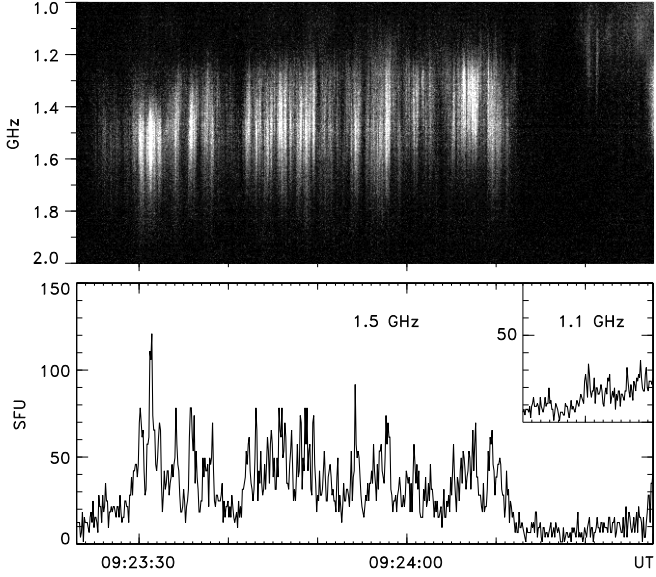}}
   % \parbox{11.5cm}{\hfill\vspace{10cm}}
  % }
  \caption{Detail of the radio observations shown in Fig.~\protect\ref{main} 
           (Ond\v{r}ejov spectrograph).}
  \label{pulsedetail}
 \end{center}
\end{figure}

Fig.~\ref{pulsedetail} displays the pulsations during the early phase of
the burst in greater detail. The intervals between the pulses and their
individual durations were not regular, leading to a quasi-periodic
appearance of the flux variations. The timescales of the pulses were lying
in the range 0.5--3~s, and the FWHM of the auto-correlation was 1.3~s. 

\begin{figure}[t]
 \begin{center}
  % \fbox{
   % \resizebox{0.84\textwidth}{!}{\includegraphics{fig/KK_threebands.ps}}
   % \resizebox{\hsize}{!}{\includegraphics{fig/KK_threebands.ps}}
   \resizebox{\hsize}{!}{\includegraphics{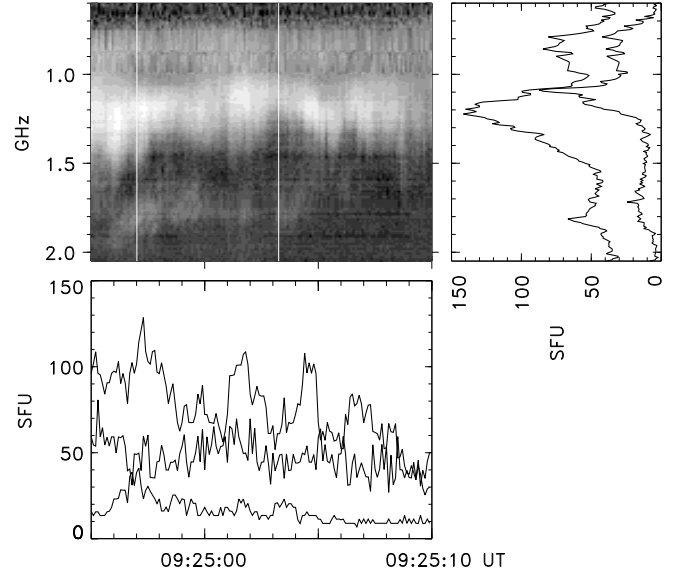}}
   % \parbox{11.5cm}{\hfill\vspace{10cm}}
  % }
  \caption{Detail of the three-band structure during part of the burst
           shown in Fig.~\protect\ref{main} (Z\"urich spectrograph). Time 
           profiles 
	   in the bottom panel are taken at 1.16, 0.90, and 1.8~GHz (in 
	   order of decreasing flux). The times of the instantaneous spectra 
	   (top right) are indicated by the white lines in the dynamic 
	   spectrum. The spectrum at 9:24:57~UT is shifted by 30~SFU 
	   for clarity.}
  \label{threebands}
 \end{center}
\end{figure}

Two additional emission bands of low intensity existed during part of the
burst (Fig.~\ref{threebands}). The high-frequency band appeared only
during a short interval (09:24:55--25:05~UT). Its drift and intensity
variations were not correlated with with the main band. The low-frequency
band, which had an average frequency drift similar to the main band and
weakly correlated intensity variations, could be discerned during
$\approx\!09$:24:20--09:25:25~UT. It is not clear whether it then decayed,
or merged with the main band, or continued to drift but was masked by the
low-frequency edge of the main band, which drifted to $\sim0.6$~GHz at
09:26:18~UT.

The HXR lightcurve does not show much structure at short timescales (most
of the fluctuations seen at the resolution of 64~ms in Fig.~\ref{main} are
noise), but the auto-correlation (e.g. in the regular interval 09:23:28 to 
09:24:13~UT) shows a FWHM of 4.5\,s, indicating significant substructure in 
time. The 25--50\,keV
BATSE lightcurve of the event considered here had been included in an
earlier search for significant timescales in hard X-ray time profiles
using a wavelet analysis method (Asch\-wanden et al.\ \cite{Asch98}), and
a weak contribution around 4~s had been found in addition to the
dominant timescale of $\sim28$~s, which reflects the overlapping main
pulses seen in Fig.~\ref{main}. 

HXR lightcurves are generally composed of a direct-precipitating component,
which carries the information on the rapid variations of the acceleration
process, and a smoothed component, leaking out of a trap, that has lost
this information. A trap model could indeed be fitted to energy dependent
time delays in the low-pass filtered HXR lightcurves of the considered
flare (Aschwanden et al.\ \cite{Asch97}).  This showed the importance of
trapping in this flare and suggested a density in the trap of the
HXR-emitting particles of order $n_{\rm tr}\sim10^{11}$~cm$^{-3}$, which
is roughly in accordance with the density of the flare loop estimated from
the soft X-ray data (see below). 

There was a high correlation between the {\sl general evolution} of the
decimetric burst and the HXR lightcurve during the initial phase: the
period of low-level emission and the impulsive onset of the main phase
were coincident. The correlation became much weaker as the decimetric flux
decreased in the course of the event. The strongest and hardest HXR peak
was accompanied by only rather weak decimetric emission, but one of the
main bunches of enhanced decimetric pulses later on ($\approx9$:25:50~UT)
was again nearly simultaneous with a significant HXR enhancement. 
The correlation of the general evolution 
shows that the decimetric burst was closely associated with the first
impulsive energy release of the flare and with the processes that led to
that energy release. Differences to the HXR burst in the later phases do
not necessarily imply a loss of association, since the radio emission in
general reflects not only the production of nonthermal particles but is
also sensitive to changes in the magnetic field, which influence the slope
of the particle distribution function and, hence, the growth rate of the
plasma waves, and to density changes in the source and along the ray path,
which influence the propagation of the radio waves. It is, however, quite
probable that the strongest and hardest HXR peak resulted from flaring in
a part of the active region that was not magnetically well connected with
the decimetric source. The situation here is typical of
comparisons between decimetric and HXR time profiles, where an association
can often be established only from the coincidence of the main features like
onset, main rise, or peak (Asch\-wanden et al.\ \cite{Asch90}).

\begin{figure}[t]
 \begin{center}
  % \fbox{
   % \resizebox{0.84\textwidth}{!}{\includegraphics{fig/KKB_cc.ps}}
   % \resizebox{\hsize}{!}{\includegraphics{fig/KKB_cc.ps}}
   \resizebox{\hsize}{!}{\includegraphics{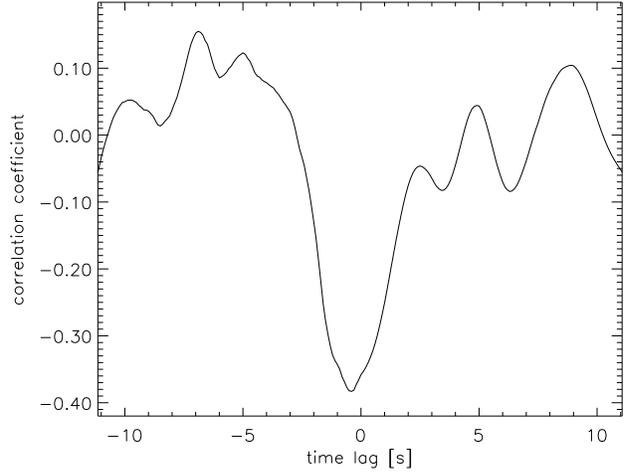}}
   % \parbox{11.5cm}{\hfill\vspace{10cm}}
  % }
  \caption{Cross correlation of the BATSE hard X-ray flux ($\ge$ 25\,keV) 
           and the decimetric flux integrated between 0.6 and 2~GHz in the 
	   interval 09:23:28--09:24:13~UT.}
  \label{cc}
 \end{center}
\end{figure}

A {\sl detailed comparison} by cross-correlation was made for the initial,
relatively regular part (shown in Fig.~\ref{pulsedetail}). The time
resolution of the BATSE data in this pre-trigger interval is 1.024\,s, so
the correlation refers to structures of at least 2\,s duration, which in 
general do not correspond to single decimetric pulses but to small
groups. The cross-correlation coefficient has a distinct minimum of $-0.38$ 
near zero lag, indicating a negative correlation (anti-correlation). Its
FWHP of only 3.0\,s suggests a trend for individual HXR pulses to peak
when the radio flux reaches a minimum. The same property appears also in
the main part of the event after subtracting a gliding-average background
with 10\,s or 20\,s time constant.

The positive correlation of the general evolution of the decimetric and
HXR lightcurves indicates the acceleration of the radiating electrons by
the same global destabilization of the magnetic field configuration, and
the anti-correlation of the details suggests acceleration by the same or
by closely related spatial substructures --- with a possibly complex
relationship between the radio and HXR sources. 

The soft X-ray images of the event, analysed by Ohyama \& Shibata
(\cite{OS98}), revealed a nearly stationary low-lying flare loop and a
plasmoid ejection, which was first seen at 9:24:40~UT at a height of
$\approx2\times10^4$~km.  In addition, a large, faint loop was apparently
connected with the rising plasmoid.  The plasmoid was continuously
accelerated and reached velocities of $\approx250$~km\,s$^{-1}$ at 9:25~UT
and $\approx500$~km\,s$^{-1}$ at 9:26~UT.  Temperatures and plasma
densities in the soft X-ray structures were estimated by Ohyama \& Shibata
(\cite{OS98}) as follows:
(a) in the flare loop at 9:24:18--9:24:50~UT: 
$T_{\rm fl}=7.1\mbox{--}18$~MK and 
$n_{\rm fl}=(0.8\mbox{--}6.7)\times10^{10}$~cm$^{-3}$ and at 
9:24:58--9:25:58~UT: 
$T_{\rm fl}=8.2\mbox{--}20$~MK and 
$n_{\rm fl}=(1.4\mbox{--}12)\times10^{10}$~cm$^{-3}$; 
(b) in the plasmoid at 9:25:00~UT:  $T_{\rm pl}=10.6$~MK and
$n_{\rm pl}=(0.8\mbox{--}1.6)\times10^{10}$~cm$^{-3}$; and 
(c) in the expanding loop at 9:26:08~UT: 
$n_{\rm el}=(3.2\mbox{--}9.4)\times10^9$~cm$^{-3}$ under the assumption of
$T_{\rm el}=2$~MK and $n_{\rm el}\sim(1.2\mbox{--}4)\times10^9$~cm$^{-3}$
for  $T_{\rm el}$ in the range 4--14~MK. 

Parameters in the current sheet, supposed to exist between the nearly
stationary flare loop and the rising plasmoid, were not explicitly given
but can be estimated from Fig.~7 of Ohyama \& Shibata (\cite{OS98}) as
$n_{\rm CS}\sim10^{10}$~cm$^{-3}$ and $T_{\rm CS}=9\times10^6$~K at
9:25:18~UT. (The latter values may actually refer to the rear side of the
plasmoid, which formed an elongated structure.)  In comparison, the most
intense radio emission in the 1.1--1.6~GHz frequency range implies source
densities in the range $n=(1.5\mbox{--}3.2)\times10^{10}$~cm$^{-3}$ in
case of fundamental or $n=(0.4\mbox{--}0.8)\times10^{10}$~cm$^{-3}$ in
case of harmonic radiation.  Although the uncertainties are substantial,
it is apparent that the densities in the radio source correspond more
closely to the densities in the plasmoid and in the current sheet than to
the densities in the flare loop or in the expanding loop (for the
reasonable assumption of $T_{\rm el}\ga4$~MK). The substantial increase of
the density in the flare loop during the event rules this loop out as site
of the radio source, since the radio emission showed, on average, a
negative frequency drift.

\section{A model based on dynamic reconnection}
\label{reconn}

Interpretations of the pulsations in the considered event in terms of
existing models for the modulation of a magnetic loop by MHD waves can be
ruled out, because these models predict pulsations with regular or slowly
varying period and most of them also with constant or exponentially
decaying amplitude (see Aschwanden \cite{Asch87}), while the observed
intervals between the pulses, the durations, and the peak intensities were
all varying rapidly and partly irregularly. 

The geometry of this {\sl eruptive} flare and the observed relation 
between HXR and decimetric radio pulsations are not suggestive of 
modulations by oscillatory nonlinear instabilities of electrons stably 
trapped for more than a minute. It appears more obvious to relate the 
pulsations to changes of
the plasma environment and to the corresponding modulations of the
particle acceleration or radio emissivity than to oscillatory nonlinear
kinetic processes in an idealized stationary trap or current sheet model. 

The Yohkoh observations provide strong evidence that a current sheet lying
above the soft X-ray flare loop and an outward propagating plasmoid were
essential elements of the flare under consideration.  The basic magnetic
topology appears to be similar to the 2D Kopp-Pneuman model of flares (see,
e.g., Tsuneta \cite{Ts96}).  Estimates of the energy content of the
plasmoid suggested that it did not contain sufficient energy to form the
current sheet by sweeping up an overlying closed magnetic field structure
(Ohyama \& Shibata \cite{OS98}).  Thus, either the plasmoid was formed and
accelerated by processes operating in the preexisting current sheet, or it
was formed jointly with the current sheet by a global MHD instability.

Considering the limited frequency extent and average frequency drift of the
observed radio burst, and the estimated densities, we interpret the burst
as emission originating from the plasmoid (or from dense plasmoid-like
structures in the current sheet).  The frequency drift of the burst can
then be attributed to the ejection of the plasmoid (or to the corresponding
upward expansion of the current sheet).

To explain the radio pulsations, we focus on small-scale MHD processes
within the current sheet, disregarding aspects of interaction with the
dense flare loop lying below the current sheet.  We expect that the
small-scale dynamics in the current sheet will be similar in the two
variants of plasmoid formation, which were mentioned above, since the
magnetic topology and the large-scale plasma flows become similar, once a
plasmoid is formed.  Here we will concentrate on the former case, i.e.,
spontaneous reconnection triggered by an initial perturbation within the
current sheet; no driving flows are prescribed at the boundaries of the
numerical system.  We note that such flows could not be specified from the
existing observations of the event.

A two-dimensional (2D) model appears to be well justified to study effects
intrinsic to the current sheet, since the size of the plasmoid along the
line of sight was estimated to be of order $10^4$~km, which exceeds the
supposed current sheet width by several orders of magnitude.  Although the
expanding loop-shaped structure, if it was really connected with the
plasmoid, implies the presence of a magnetic guide field component, we
restrict ourselves here further to the case of vanishing guide field,
i.e., an initially antiparallel (``neutral'') current sheet is studied. 
As long as the internal current sheet dynamics including the formation and
acceleration of the plasmoid are concerned, this simplification also
appears reasonable within the 2D approximation for the following reason. 
In 2D incompressible MHD, an initially homogeneous guide field is only
passively advected, it has no influence on the dynamics.  In 2D
simulations of compressible current sheet dynamics, its influence has been
found to remain very small (Ugai \cite{Ug93}).  (A guide field becomes
essential in 3D reconnection studies, where it can lead to the formation
of structures like interlinked flux tubes, which are otherwise absent.)

Magnetic reconnection is often considered in the framework of the
stationary Petschek model (e.g., Ugai \cite{Ug92}, \cite{Ug95}; Yokoyama
\& Shibata \cite{YS94}; Tsuneta \cite{Ts96}) and its generalizations
(Priest \& Forbes \cite{PF86}).  However, there is a long-lasting dispute
whether stationary Petschek reconnection is possible (Biskamp \cite{Bi86};
Forbes \& Priest \cite{FP87}; Scholer \cite{Scho91}; Kulsrud
\cite{Ku98}).  For long current sheets, numerical simulations indicated
that resistive magnetic reconnection proceeds in a time-variable manner,
involving repeated formation of magnetic islands and their subsequent
coalescence (Biskamp \cite{Bi82,Bi94}; Scholer \& Roth \cite{SR87}; Karpen
et al.\ \cite{KAD95}).  This is particularly true if the resistivity is
permitted to vary.  The process is known as ``secondary tearing'' or as
the ``impulsive bursty'' regime of reconnection (Priest \cite{Pr85}).  The
repeated coalescence of islands may lead to the formation of a growing,
and eventually large-scale, plasmoid (Scholer \& Roth \cite{SR87};
Schumacher \& Kliem \cite{SK96}; Magara et al.\ \cite{Ma97}). Coincident
with the formation of each new magnetic island, the electric field at the
new magnetic X point is enhanced.  This probably leads to a burst of
accelerated particles (although the detailed mechanisms of particle
acceleration in reconnection regions still remain unknown).  In the
simplest model, each formation of a new magnetic X point corresponds to a
burst of acceleration, which, in its turn, corresponds to an enhancement
of the radio flux (one pulsation).  More complicated schemes are
conceivable and briefly discussed in Sect.~\ref{discuss}.

\subsection{Basic equations and simulation method}

The equations of compressible, resistive one-fluid MHD used in this paper are
%-------------------------------------------------------------------------
\begin{eqnarray}
\partial_{t}\rho&=&
                 -\nabla\cdot(\rho\,{\bf u}),               \label{eq_rho}\\
\rho\,\partial_{t}{\bf u}&=&
                 -\rho\,(\,{\bf u}\cdot\nabla\,)\,{\bf u}
                 -\nabla\,p+{\bf j}\mbox{\boldmath$\times$}{\bf B},\,\,\,\,\\
\partial_{t}{\bf B}&=&
    \nabla\mbox{\boldmath$\times$}(\,{\bf u\mbox{\boldmath$\times$} B}\,)
   -\nabla\mbox{\boldmath$\times$}(\,\eta\,{\bf j}\,),\\
\partial_{t}U&=&
              -\nabla\cdot{\bf S},                         \label{eq_energy}
\end{eqnarray}
%-------------------------------------------------------------------------
\noindent
where the current density {\bf j}, the total energy density $U$, and the flux
vector ${\bf S}$ are given by
%-------------------------------------------------------------------------
\begin{eqnarray}
{\bf j}&=& \frac{1}{\mu_0}\,\nabla\mbox{\boldmath$\times$}{\bf B},
\nonumber\\
U      &=& \rho\,w+\frac{\rho}{2}\,u^{2}+\frac{B^{2}}{2\,\mu_0},
\nonumber\\
{\bf S}&=&(\,U+p+\frac{B^{2}}{2\,\mu_0}\,){\bf u}
         -({\bf u}\cdot{\bf B})\frac{{\bf B}}{\mu_0}
         +\eta{\bf j}\mbox{\boldmath$\times$}\frac{{\bf B}}{\mu_0},
\nonumber
\end{eqnarray}
%-------------------------------------------------------------------------
and $w$ is the internal energy per unit mass, which is related to the
pressure through the equation of state, $p=(\,\gamma-1\,)\,\rho\,w$.  In
the two-dimensional model adopted here, $\partial/\partial z\!=\!0$ for
all quantities and $B_z=u_z=0$; hence the current density and the electric
field possess only $z$ components.  Further, the ratio of specific heats
is $\gamma\!=\!(N+2)/N\!=\!2$, where $N$ is the number of degrees of
freedom.  The electric field is given by 
%-------------------------------------------------------------------------
\begin{eqnarray}
{\bf E}=-{\bf u}\mbox{\boldmath$\times$}{\bf B}+\eta\,{\bf j}.
\label{Ohm}
\end{eqnarray}
%-------------------------------------------------------------------------

An antiparallel Harris equilibrium (a neutral current sheet) with uniform
density is chosen as the initial condition:
%----------------------------------------------------------------------------
\begin{eqnarray}
{\bf B} &=&-\,B_0\,\tanh(y/l_{\rm CS})\,{\bf e}_x\,,           \label{eq1}\\
{\bf u} &=& 0\,,                                                          \\
\rho    &=& \rho_0\,,                                                     \\
p       &=& (1+\beta)B_0^2/(2\mu_0)-B_x^2/(2\mu_0)\,,          \label{pre}
\end{eqnarray}
%---------------------------------------------------------------------------
where the plasma beta is defined as $\beta=2\mu_0\,p(|y|\to\infty)/B_0^2$.

The variables are normalized by quantities derived from the current sheet
half width $l_{\rm CS}$ and the asymptotic ($|y|\to\infty$) Alfv\'en
velocity $V_A=B_0/(\mu_0\rho_0)^{1/2}$ of the configuration at $t=0$. 
Time is measured in units of the Alfv\'en time $\tau_A=l_{\rm CS}/V_A$,
and $p$, {\bf E}, {\bf j}, and $\eta$ are normalized by $B_0^2/(2\mu_0)$,
$V_AB_0$, $B_0/(\mu_0l_{\rm CS})$, and $\mu_0l_{\rm CS}V_A$,
respectively.  Then the Lundquist number is given by $S=\eta^{-1}$.  The
normalized variables will be used in the remaining part of this subsection.

The equilibrium is initially perturbed by a localized resistivity
%----------------------------------------------------------------------------
\begin{equation}
\eta(x,y,t\le t_0)\!=\!0.02\exp[-(x/0.8)^2-(y/0.8)^2]\,,
\end{equation}
%----------------------------------------------------------------------------
fixed for a few Alfv\'en times, $t_0=10$.  Thereafter the resistivity is
variable and is determined self-consistently at every grid point and time
step from the electron-ion drift velocity $v_{\rm D}=v_0\,j/\rho$ (where
the factor $v_0=v_{\rm D}(0)=d_i/l_{\rm CS}$ results from the
normalization and $d_i=c/\omega_{pi}$ is the ion inertial length).  An
``anomalous'' value, $\eta_{\rm an}$, is set if a threshold, $v_{\rm cr}$, is
exceeded, otherwise vanishing resistivity is assumed:
%----------------------------------------------------------------------------
\begin{equation}
\eta({\bf x},t)=\left\{
  \begin{array}{l@{\quad:\quad}l}       0       & |v_{\rm D}|\le v_{\rm cr}\\
    \begin{displaystyle}
     C\,\frac{(|v_{\rm D}({\bf x},t)|-v_{\rm cr})}{v_0}
    \end{displaystyle}                          & |v_{\rm D}| >  v_{\rm cr}.
  \end{array}
 \right.
\label{etaandef}
\end{equation}
%----------------------------------------------------------------------------
Here $C\!=\!0.003$ and $v_{\rm cr}=3v_0$.  Such a resistivity model is
customary in MHD simulations of magnetic reconnection (e.g., Ugai
\cite{Ug92}; Yokoyama \& Shibata \cite{YS94}).  We have verified that the
dynamics of the system does not significantly depend on the particular
choice of the parameters $t_0$, $C$, and $v_{\rm cr}$ in a substantial
range around the values used here.  Also the dependence on $\beta$ is
weak.  Simulations with $\beta>1$ show less variable secondary tearing. 
We present a low-beta simulation, $\beta=0.15$.

Eqs.~(\ref{eq_rho})-(\ref{eq_energy}) are transformed into a flux conserving
form,
$\partial_t{\bf\Psi}+\partial_x{\bf F}({\bf\Psi})
                    +\partial_y{\bf G}({\bf\Psi})=0$,
for each of the six integration variables,
${\bf \Psi}=(\rho,\rho\,u_{x},\rho\,u_{y},B_{x},B_{y},U)$, where {\bf F} and
{\bf G} are the nonlinear flux terms.  A two-step Lax-Wendroff scheme
according to Ugai \& Tsuda (\cite{UT77}) and a uniform Cartesian grid with
resolutions $\Delta x=\Delta y=0.045$ are employed for the numerical
integration.  In order to stabilize the scheme, artificial smoothing (Sato \&
Hayashi \cite{SH79}) is applied such that the variables at each grid point are
replaced after each full time step of the Lax-Wendroff algorithm in the
following manner
%-------------------------------------------------------------------------
\begin{displaymath}
{\bf\Psi}_{ij}^{n}\longrightarrow
     \lambda{\bf\Psi}_{i,j}^{n}+\frac{1\!-\!\lambda}{4}
     ({\bf\Psi}_{i+1,j}^{n}\!+\!{\bf\Psi}_{i-1,j}^{n}\!+\!
      {\bf\Psi}_{i,j+1}^{n}\!+\!{\bf\Psi}_{i,j-1}^{n})
\end{displaymath}
%-------------------------------------------------------------------------
with $\lambda$ varying in the range 0.97--0.98 during the simulation. 

In order to study the effect of secondary tearing, the numerical box must not
be too small.  For box lengths $L_x\sim20$, which is a common value in the
literature, secondary tearing occurs only for rather small box heights,
$L_y\sim4$.  Increasing the box length to $L_x\sim40$, leads to secondary
tearing for all reasonable values of $L_y$, however, the details of the time
profile of the reconnection rate, in particular the development of irregular
or quasi-periodic reconnection, depend on $L_x$ until $L_x$ exceeds a value
$\sim60$.  The simulation presented here used $L_x=64$ and $L_y=16$.

The simulation is carried out in the first quadrant, using symmetrical
boundary conditions at the lower and left boundaries.  Open boundary
conditions are realized at \{$x\!=\!L_x$\} and \{$y\!=\!L_y$\} by the 
requirement
that the normal derivatives of all variables vanish, except for the normal
component of ${\bf B}$, which is determined from the solenoidal condition.

\subsection{Dynamic magnetic reconnection}
\label{dynamic}

\begin{figure*}
 \begin{center}
   % \fbox{
    % \resizebox{0.84\textwidth}{!}{\includegraphics{fig/KK_Bej_evol.ps}}
    % \resizebox{\hsize}{!}{\includegraphics{fig/KK_Bej_evol.ps}}
    \resizebox{\hsize}{!}{\includegraphics{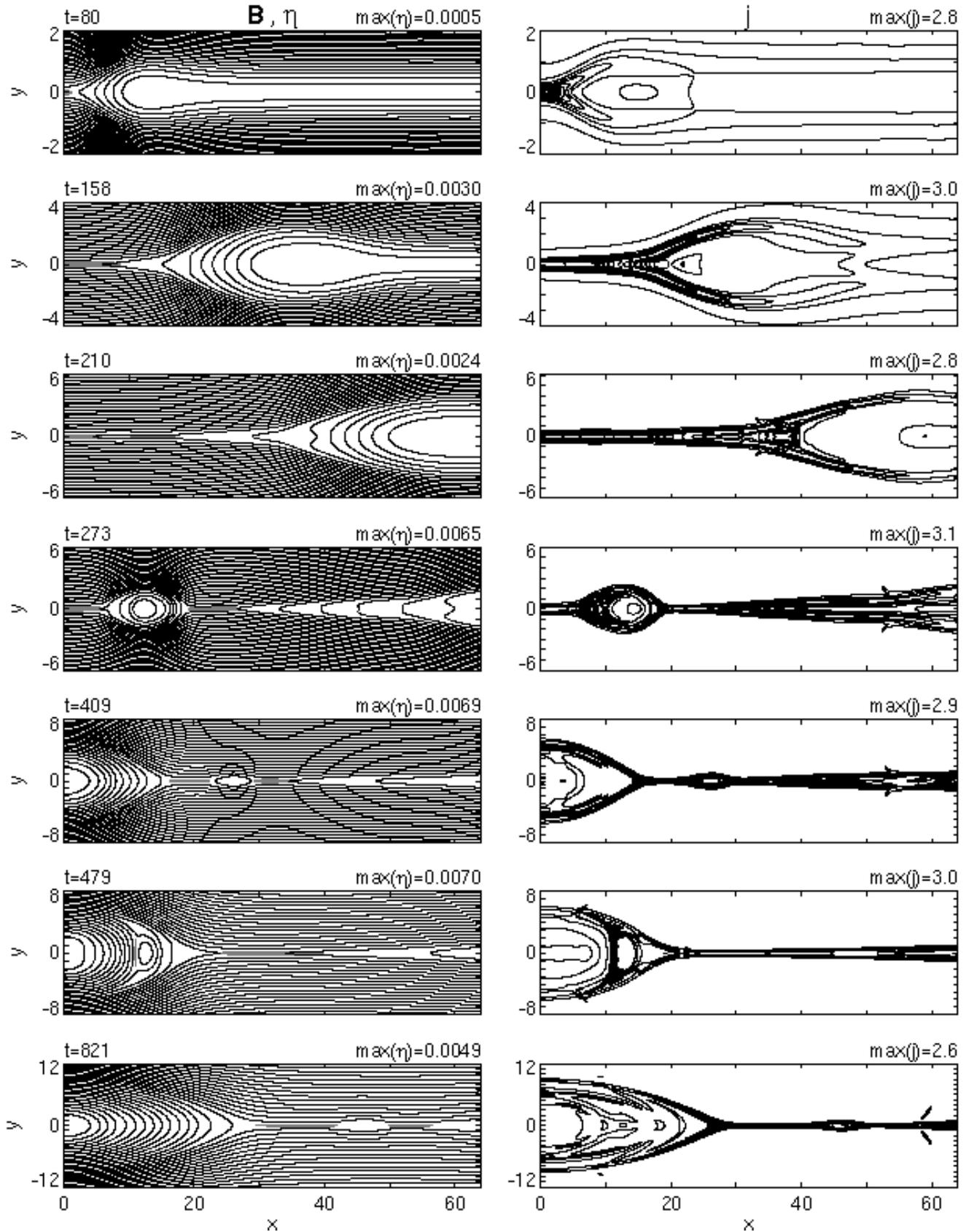}}
    % \parbox{17cm}{\hfill\vspace{21cm}}
   % }
  \caption{Magnetic field (left panels) and current density (right panels) at
           characteristic times of the evolution. The regions, where
           anomalous resistivity is excited, are shown shaded in the magnetic
           field plots.}
  \label{Bej_evol}
 \end{center}
\end{figure*}

\begin{figure*}
 \begin{center}
   % \fbox{
    % \resizebox{0.84\textwidth}{!}{\includegraphics{fig/KK_upr_evol.ps}}
    % \resizebox{\hsize}{!}{\includegraphics{fig/KK_upr_evol.ps}}
    \resizebox{\hsize}{!}{\includegraphics{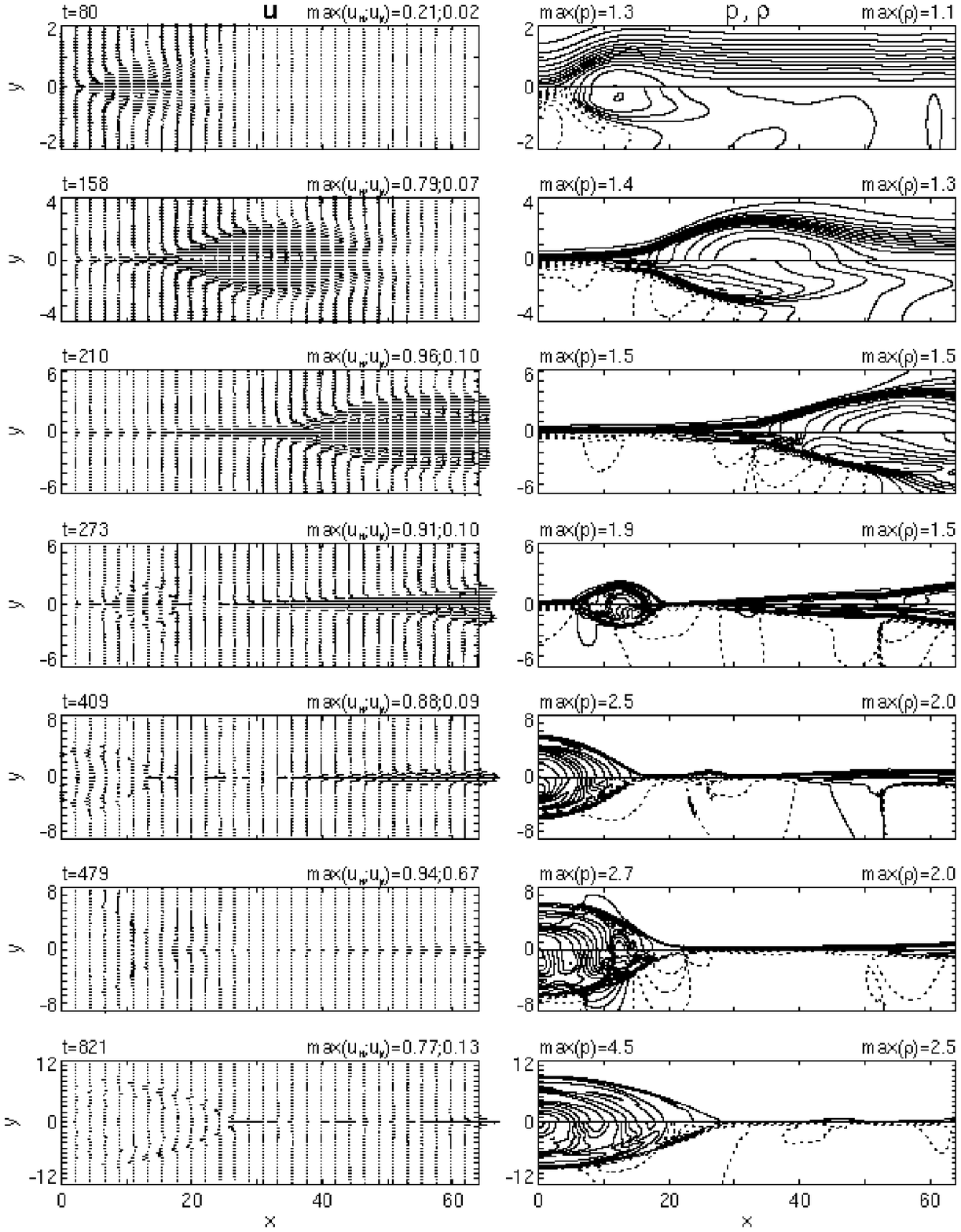}}
    % \parbox{17cm}{\hfill\vspace{21cm}}
   % }
  \caption{Velocity field (left panels), pressure (right panels, $y>0$), and
           density (right panels, $y<0$) at the same times as in
           Fig.~\protect\ref{Bej_evol}. Contours of rarefactions ($\rho<1$) 
           are shown dashed.}
  \label{upr_evol}
 \end{center}
\end{figure*}

The dynamical evolution of the considered system, shown in
Figs.~\ref{Bej_evol} and \ref{upr_evol}, starts as in many other
investigations of spontaneous reconnection (e.g., Ugai \cite{Ug92}).  The
initial perturbation causes a small burst of reconnection at the origin,
which leads to acceleration of plasma along the $x$ axis by the Lorentz
force $f_x=-jB_y$.  Mass conservation then enforces an inflow $u_y$ into
the region of the initial perturbation.  This convects new magnetic flux
toward the sheet and causes the current density to increase again
locally.  After $\sim70\,\tau_A$, the threshold $v_{\rm cr}$ is exceeded
and anomalous resistivity is spontaneously excited at the origin.  This
enables a positive feedback between reconnection, acceleration of the
fluid, and the local rise of the resistivity.  The continuing and
initially amplifying reconnection leads to a strong outflow of heated
plasma along the $x$ axis, which eventually reaches the Alfv\'en velocity
and drives a high-pressure region in front of it toward the outer
boundary.  The state of the system during this phase of amplifying
Petschek-like reconnection is illustrated in Figs.~\ref{Bej_evol} and
\ref{upr_evol} at $t=80$ and $t=158$, where the current density ridges
near the separatrix lines of the magnetic field delineate the slow mode
shocks.

Since the level of anomalous resistivity at the origin can adjust to
changes in the outer region, it has often been assumed that a balance
between convection and diffusion is reached and that, consequently, the
Petschek-like reconnection state, once formed, does approach stationarity
(e.g., Ugai \cite{Ug92}).  Also many observational studies of solar
eruptions have assumed that stationary Petschek reconnection occurred
(e.g., Tsuneta \cite{Ts96}).  However, stationary Petschek reconnection
has been obtained in numerical experiments only if a localized resistivity
was held fixed in time or if the box length was small, $L_x\la20$ (see
Scholer \cite{Scho91}).  Variable resistivity leads to non-stationary
reconnection if the current sheet is sufficiently extended, independent of
whether the resistivity model is based on the electron-ion drift velocity
or on the current density (Scholer \& Roth \cite{SR87}).  This has
recently been confirmed for 3D reconnection as well (Schumacher et al.\
\cite{SKS00}).

The principal effect that leads to variable reconnection appears to be the
inability of the fluid to carry a sufficient amount of magnetic flux into
the strongly localized diffusion region to support the Alfv\'enic outflow
of reconnected flux in steady state (Kulsrud \cite{Ku98}).  The free
acceleration of the plasma, which is penetrated by reconnected field lines,
into the $x$ direction causes the inflow to steepen the current density
around the origin in a range of increasing extent.  Although, according to
Eq.~(\ref{etaandef}), the density depression at the origin tends to keep
the anomalous resistivity more localized than the range of enhanced current
density, also the area of anomalous resistivity excitation increases into
the $x$ direction.  Its extent reaches $\delta x_\eta=13$ at $t=200$.  At
the same time, the width of the anomalous resistivity region in the $y$
direction stays approximately constant, $\delta y_\eta\approx0.2$.  The
central part of the current sheet has thus formed a Sweet-Parker
configuration, which implies a reduction of the reconnection rate in
comparison to the Petschek configuration ($t\approx100\mbox{--}200$).  The
acceleration of the fluid is then also reduced locally, and both the
current density peak and the density depression at the origin are reduced.
A resistivity maximum remains at the edge of the elongated $\eta_{\rm an}$
area, while the resistivity drops to zero between the new maximum and the
origin.  This leads immediately to the formation of a new X point (and its
mirror image) at $t\approx200$ --- the so-called secondary tearing (Scholer
\& Roth \cite{SR87}).

The sequence then repeats at the new X point, where the level of the
resistivity and the field line reconnection are initially also strongly
amplified.  This drives the coalescence of the magnetic islands between the
three X points, resulting in a plasmoid at the origin.  Over the next
$\sim200$ Alfv\'en times, the magnetic structure at the new X point becomes
flat and the resistivity area elongates and splits again, forming a new X
point further out ($t\approx400$).  The corresponding magnetic island is again
driven toward the central plasmoid by the rapidly evolving new X point.  Over
a period of $1200\,\tau_A$, seven strong new X points are created,
supplemented by a number of weak X and O pairs which can be considered as
``fluctuations'' and are of little dynamical importance.  All seven magnetic
islands merge with the central plasmoid, while some of the ``fluctuations''
merge with the outflow.  This merging is accompanied by transient excitations
of anomalous resistivity between the approaching islands (see
Fig.~\ref{Bej_evol} at $t=479$).

\begin{figure}
 \begin{center}
  % \fbox{
   % \resizebox{\hsize}{!}{\includegraphics{fig/KK_R_rate.ps}}
   \resizebox{\hsize}{!}{\includegraphics{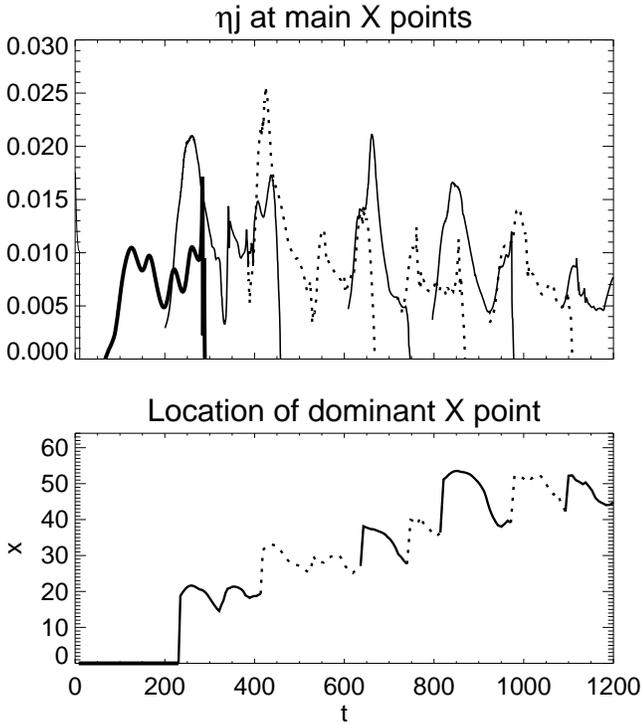}}
   % \parbox{8.5cm}{\hfill\vspace{9.5cm}}
  % }
  \caption{Top: Electric field at the main X points (reconnection rate) 
                vs.\ time.
            Thick line: Petschek-like reconnection with a central X point 
                ($t\approx70\mbox{--}200$) and forced reconnection by 
                initial island coalescence at the origin 
                ($t\approx200\mbox{--}300$). 
            Medium line width (solid and dotted): main new X points.
            Thin line: initial perturbation.
           Bottom: Location of the dominant X point.}
  \label{R_rate}
 \end{center}
\end{figure}

An integral view of the dynamics is given in Fig.~\ref{R_rate}, where the
electric field at the various main X points (i.e., the reconnection rate)
and the location of the dominant X point in the system are plotted.  The
seven new main X points are separated using different linestyles.  Their
individual variations contain imprints of the additional formations of weak
X points, which disturb the magnetic configuration and the outflow from the
main X points.  The ``dominant'' X point in the bottom panel is the one
where instantaneously the highest reconnection rate occurs.  The
change-over of maximum reconnection to a stronger new X point further out
marks jumps in the position of the dominant X point.  It is apparent from
the figure that the secondary tearing leads to higher reconnection rates
than the Petschek-like reconnection, which exists only during the first
$\sim200\,\tau_A$.  The reconnection rate in the numerical experiment is of
the same order as that implied by the observations, $R_{\rm obs}\sim0.02$
(Ohyama \& Shibata \cite{OS98}).  The intervals between the reconnection
peaks are not uniform, but also not completely irregular.

As a result of secondary tearing, the central plasmoid is continuously fed
with heated material carrying reconnected magnetic flux.  The highly
variable inflow into the plasmoid from the neighbouring X points triggers
oscillations of the plasmoid as a whole.  These oscillations in their turn
enhance the time variability of the highly nonlinear, threshold-depending
reconnection process in the outer parts of the extended current sheet. 
The pressure and density in the plasmoid are continuously rising, causing
it to expand in the $x$ and $y$ directions.  This expansion appears to be
the reason of the general slow decline of the activity seen in
Fig.~\ref{R_rate} after $t\approx400$.  The increasing width of the
plasmoid in the $y$ direction makes the convection of new magnetic flux
toward the neighbouring X point more difficult because an increasing
amount of field line bending is required, and the space available to form
an extended flat current sheet section between the plasmoid and the
boundary $L_x$ is reduced.  This decrease of the average reconnection rate
is expected to be far smaller in the solar corona, where the magnetic islands
are much more elongated in the $x$ direction and less field line bending
is required (see Sect.~\ref{scales}). 

\begin{figure}
 \begin{center}
  % \fbox{
   % \resizebox{\hsize}{!}{\includegraphics{fig/KK_plasmoid.ps}}
   \resizebox{\hsize}{!}{\includegraphics{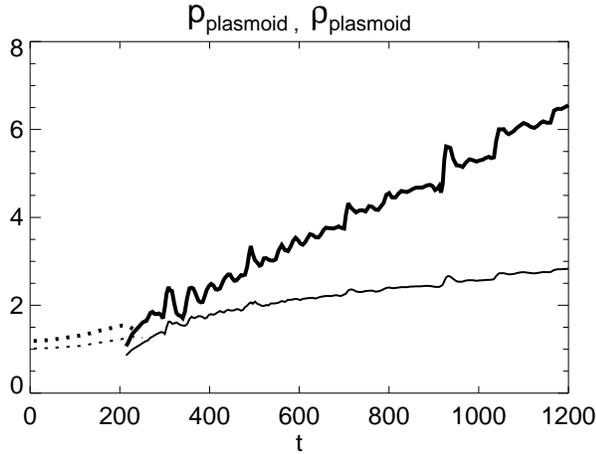}}
   % \parbox{8.5cm}{\hfill\vspace{6cm}}
  % }
  \caption{Peak pressure (thick lines) and density (thin lines) of the
           plasmoid formed by the initial perturbation (dotted) and of the
           central plasmoid formed by secondary tearing.
           The initial normalized values are $p_0(y\!=\!0)=1+\beta=1.15$ and
           $\rho_0=1$.}
  \label{plasmoid}
 \end{center}
\end{figure}

We have seen that the dynamical evolution leads to two types of dense
structures.  The first one is the symmetric pair of high-pressure regions
driven ahead of the outflow, which commences with the initial perturbation.
This structure is commonly referred to as plasmoid, although only the field
lines at its rear side are closed, while the field lines at its front side
close at infinity.  (In the numerical experiment, a small fraction of the
flux closes also at its front side, due to the nonvanishing numerical
diffusion.)  The second one is the plasmoid resulting from secondary
tearing.  Fig.~\ref{plasmoid} shows that pressure and density in both
plasmoids rise with time to values far in excess of the initial ones and
that the rise is much stronger in the central plasmoid.  For both plasmoids
one can see that the density ratio $r_n=\rho_{\rm pl}/\rho_0$ remains
much smaller than the corresponding pressure ratio.  This is due to the
strong plasma heating in the dissipation region around the X points.  In
the solar corona one can expect that the size ratio of dissipation region
and plasmoid is far smaller than in the simulation and that the pressure
enhancement is to a larger fraction realized as a density enhancement.  The
soft X-ray images of the plasmoid in the considered flare suggest a density
ratio to the surrounding corona of $\sim10$ (Ohyama \& Shibata
\cite{OS98}).  Keeping in mind that the overall scales of the numerical
experiment, $L_x=64$ and $t_{\rm max}=1200$, are by orders of magnitude
smaller than those of the flare, the formation of a plasmoid with a
density enhancement $r_n\sim10$ in the flare by the secondary tearing and
subsequent coalescence processes appears to be possible.

The central plasmoid cannot move in the symmetrical simulation. In a small
range of parameter values, it can tear and leave the box in two parts
(Kliem \& Schumacher \cite{KS96}).  If the condition of symmetry about the
$y$ axis is relaxed, small asymmetries build up, which eventually lead to
strong acceleration of the plasmoid.  Previous simulations showed that
a plasmoid can be formed also in the presence of asymmetries if their
initial value is small (Schumacher \& Kliem \cite{SK96}).  There it was
further found that the pressure and density in the plasmoid continue to
rise during the acceleration and ejection of the plasmoid and that the
velocity of the plasmoid can reach a significant fraction of the Alfv\'en
velocity ($\sim0.3\,V_A$). 

As long as the plasmoid is being formed at a fixed position, its density
is increasing. Only if the plasmoid starts to move along the current sheet
to greater heights, it will expand and its pressure and density will
decrease again according to the decreasing pressure of the surrounding
medium. The density {\sl ratio} to the surrounding material may stay
constant, or may even increase further, as long as the reconnection
continues.

\subsection{Particle acceleration, trapping, and radio emission}

It is well known that particles are accelerated near the X points in the
DC electric field associated with magnetic reconnection (e.g., Deeg et al.\
\cite{DBD91}; Moses et al.\ \cite{MFL93}; Birn et al.\ \cite{Bi98}).  A
multiple X point reconnection process, characteristic of secondary tearing
and other forms of impulsive bursty reconnection (like the multiple
coalescence instability), strongly enhances the number of accelerated
particles and their maximum energy in comparison to single X point
reconnection, where the particles can quickly escape (Ambrosiano et al.\
\cite{Am88}; Scho\-ler \& Jamitzky \cite{SJ87}; Kliem \cite{Kl94}; Kliem
et al.\ \cite{KSS98}).  Peaks of the reconnection rate, such as seen in
Fig.~\ref{R_rate}, represent temporal peaks of the electric field at the X
points and, hence, correspond to peaks of particle acceleration by the DC
electric field.  The same is true for models of stochastic particle
acceleration which suppose that the primary MHD wave spectrum is excited
by the reconnection process (Miller et al.\ \cite{Mi96}). 
Although the physics of particle
acceleration in reconnection regions still appears to be far from being
understood, the existing investigations strongly suggest that a
significant fraction of the particles in a dynamic current sheet can be
accelerated to near-relativistic and relativistic energies.  An
investigation of particle acceleration in a current sheet that undergoes
secondary tearing is planned for a future study.

The strongest electric fields occur at the main magnetic X
points adjacent to the plasmoid and, thus, a large fraction of the
accelerated particles becomes trapped in the plasmoid.  The other part can
escape along the field lines of the reconnection outflow.  Those particles
escaping with the downward directed outflow are likely to become trapped
in the magnetic cusp configuration between the current sheet and the
underlying flare loop.  Such trapping may explain the dominance of the
trap-leakage component in the hard X-ray lightcurve and contribute to the 
decrease of the correlation between the general evolution of the hard
X-ray and decimetric flux profiles in the course of the event (i.e., with
increasing plasmoid height). 

In a strictly two-dimensional configuration, the particles in the central
plasmoid are trapped indefinitely (e.g., Scholer \& Jamitzky
\cite{SJ87}).  In reality, also this part can escape from the plasmoid
along the $z$ direction, either by (slow) guiding centre drifts into the
$z$ direction, or by free streaming if a guide field $B_z\ne0$ exists. 

It is not clear whether the dynamic current sheet that forms a plasmoid
does also form a loop-shaped magnetic trap (which is the canonical source
model of several radio burst types, including decimetric continuum
bursts).  A highly distorted magnetic loop, or flux rope, may be
associated with the central plasmoid if a significant guide field
component ($B_z>B_{x0}$) is present.  Such a configuration was, in fact,
suggested for the considered flare by Ohyama \& Shibata (\cite{OS98}). 
However, due to the magnetic shear introduced by a $B_z\ne0$ in the
vicinity of the plasmoid, the photospheric footpoints of the field lines
that are connected to the plasmoid may be widely dispersed in this case,
and it is not clear whether a loss-cone distribution function can be
formed.  Therefore, it is possible that an alternative source model is
required for the radio burst.  Since the accelerated particles may gain
excess perpendicular energy from drifts in the inhomogeneous magnetic
field near the X points (Kliem \cite{Kl94}), they may be immediately
unstable (i.e., within the current sheet and the plasmoid) to excitation
of upper hybrid waves, which easily transform into radio emission, and the
presence of a magnetic trap may not be necessary to form a radio source.

\subsection{Scales in coronal current sheets}
\label{scales}

In this section we investigate how the scales found in the numerical
experiment transform for parameters of the middle solar corona, where soft
X-ray ejecta and decimetric radio sources are generally formed.  As
characteristic plasma parameters in the centre of the current sheet, we
take the values that can be inferred for the flare of October 5, 1992 from
Figs.~6 and 7 in Ohyama \& Shibata (\cite{OS98}), i.e.,
$n_{\rm CS}\sim10^{10}$~cm$^{-3}$ and $T_{\rm CS}=9\times10^6$~K
(Sect.~\ref{obs}).  These values refer to a dense and heated (i.e.,
presumably active) current sheet or possibly to the trailing part of the
extended plasmoid, which had not yet reached a great height by the time of
these parameter determinations (9:25:00--9:25:18~UT).  The plasma
surrounding the current sheet was much less dense and much less hot, as was
the current sheet itself in the Yohkoh/SXT images at a later stage of the
evolution.  A typical value of the coronal background plasma density at
the considered heights ($h\sim2\times10^9$~cm) is $n_0\sim10^9$~cm$^{-3}$.
As coronal background temperature we will take $T_0\sim2.5\times10^6$~K.
The maximum instantaneous bandwidth of the decimetric radio burst suggests
that the density varied by a factor $\approx4$ within the radio source.

Although the plasma parameters estimated by Ohyama \& Shibata are subject
to uncertainties that derive from necessary assumptions on the scale of
the structures along the line of sight ($L_z=10^9$~cm) and the filling factor
($f=1$), they are consistent with the parameters derived for another flare
which showed a plasma ejection (Tsuneta \cite{Ts97}).  We note that a
density of $\sim10^{10}$~cm$^{-3}$ at the electron acceleration site was
found also for other flares, using a different method (by determining the
starting frequency of forward and reversely drifting type III burst pairs;
see Aschwanden et al.\ \cite{Asch93}).

Estimates of the magnetic field strength $B_0$ in the plasma surrounding
the current sheet are also rather uncertain. Ohyama \& Shibata
(\cite{OS98}) obtained $B_0\sim70$~G from the condition of pressure 
equilibrium across the current sheet,  $B_0^2/8\pi=P_{\rm CS}-P_0$, using
a rough estimate of the internal and external pressures, $P_{\rm CS}$ and
$P_0$, respectively (actually they supposed a higher pressure than implied
by the values of $n_{\rm CS}$ and $T_{\rm CS}$ given above). This way of
estimating $B_0$ is rather sensitive to the filling factor of the
high-pressure current sheet plasma in the area of one detector pixel, 
$B_0\propto f^{-1/2}$ for $P_{\rm CS}\gg P_0$. Hence, a smaller as well as
a higher value of the magnetic field would also be plausible. 
Nevertheless, we shall use $B_0\sim70$~G, which gave a reasonable estimate
of the flare energy balance.  Using this  with $n_0$ and $T_0$, the
Alfv\'en velocity and the plasma beta outside of the current sheet become
$V_A\sim4.8\times10^8$~cm\,s$^{-1}$ and $\beta\sim3.5\times10^{-3}$,
respectively. 

The current sheet half width can be estimated from the threshold condition
for onset of kinetic current-driven plasma instabilities.  Spontaneous
excitation and re-excitation of anomalous resistivity implies that the
ion-electron drift velocity in the sheet remains close to the critical
value $v_{\rm cr}$.  This critical drift velocity is of the order of the ion
thermal velocity, $v_{ti}=(T_i/m_i)^{1/2}$, for the three most relevant
instabilities and a wide range of parameter values.  The three most
relevant instabilities are the lower-hybrid-drift instability, the
electrostatic ion cyclotron instability, and the ion sound instability
(see, e.g., the review by Papadopoulos \cite{Pa79}).  Using
$v_{\rm cr}=v_{ti}$, Ampere's law yields $l_{\rm CS}\sim4\beta^{-1}r_{ci}$, 
where $r_{ci}=v_{ti}/\omega_{ci}$ is the ion cyclotron radius.  For our
external values $n_0$ and $T_0$ we find $l_{\rm CS}\sim2\times10^4$~cm. 
The corresponding Alfv\'en time is only $5\times10^{-5}$~s.

When considering the direction along the magnetic field, we have to recall
that a fluid model is valid only on scales $L$ larger than the mean free
path, $L>\lambda_{\rm mfp}$.  In the fluid model, this length is the lower
limit of inhomogeneity scales of $\eta$, which set the distance $\delta_x$
between neighbouring X points.  Since the formation of inhomogeneities in
the current sheet is permanently enforced by the outflow along the $x$
axis of material that carries reconnected field lines, it appears
reasonable to suppose that $\delta_x\sim\lambda_{\rm mfp}$.  The mean free
path is of order $5\times10^7$~cm for both pairs of density and
temperature values $(n_0,\,T_0)$ and $(n_{\rm CS},\,T_{\rm CS})$ [but
varies by a factor $\sim10^{\pm1}$ for the other combinations
$(n_0,\,T_{\rm CS})$ and $(n_{\rm CS},\,T_0)$].

In the simulation, the time interval between main peaks of the
reconnection rate is of order $t_R\sim200\,\tau_A$, which is about
$\sim13\,\delta_x/V_A$ (Alfv\'en crossing times based on the average
distance ($\delta_x\approx15\,l_{\rm CS}$) between neighbouring X
points).  We now scale the timescale $t_R\sim13\,\delta_x/V_A$ to the much
larger distances expected in the corona, using 
$\delta_x\sim\lambda_{\rm mfp}$ (which implies 
$t_R\propto n_0^{-1/2}\,T_0^2\,B_0^{-1}$). 
For $B_0=70$~G and three different values of the density, we
derive the timescale $t_R$ in the range of temperatures
$T=(2.5\mbox{--}9)\times10^6$~K.  This yields $t_R=1\mbox{--}13$~s for
$n=10^9$~cm$^{-3}$; $t_R=0.7\mbox{--}8$~s for $n=3\times10^9$~cm$^{-3}$;
and $t_R=0.4\mbox{--}4$~s for $n=10^{10}$~cm$^{-3}$.  Since the elongation
and structuring of the area of anomalous resistivity (which is the
dominant factor controlling the dynamics) takes place in the inflow
sections of the current sheet, where the density and temperature are close
to the external values $n_0$ and $T_0$, the estimate $t_R\sim1$~s may be
the most realistic one.  In view of the uncertainties of the parameter
determinations, this is amazingly close to the observed intervals between 
the radio pulses, $t_{\rm radio}\approx0.5\mbox{--}3$~s.

The Poynting flux of electromagnetic energy toward the current
sheet is ${\cal P}_{\rm in}=2(B_0^2/\mu_0)\bar{u}_{\rm in}L_{\rm CS}L_z$, 
where $L_{\rm CS}$ is the vertical ($x$-) extent of the current sheet and
the average inflow velocity can be expressed using the average
reconnection rate $\bar{R}$ as $\bar{u}_{\rm in}\sim\bar{R}V_A$. Using 
$L_{\rm CS}\sim L_z\sim10^9$~cm and $\bar{R}\sim0.02$ (from
Fig.~\ref{R_rate}), we find  
${\cal P}_{\rm in}\sim8\times10^{27}$~erg\,s$^{-1}$. This appears to be a
reasonable value, since it exceeds the increase rate of the thermal energy
in the flare loop obtained by Ohyama \& Shibata (\cite{OS98}) by a
moderate factor ($\sim3$) consistent with the fact that not all of the
inflowing electromagnetic energy can be dissipated. The scaling 
${\cal P}_{\rm in}\propto B_0^3\,n_0^{-1/2}$ shows that higher energy
inflow rates can easily be achieved by modest increases of $B_0$. 

On the other hand, the rate of particles flowing into the current sheet,
$\dot{N}=2n_0\bar{u}_{\rm in}L_{\rm CS}L_z$, seems to fall short of the
requirement on the acceleration rate of HXR-emitting particles, which is
on the order of $\sim10^{35}\mbox{--}10^{36}$~s$^{-1}$ for an M class
flare.  Our parameters $n_0$, $B_0$ yield
$\dot{N}\sim2\times10^{34}$~s$^{-1}$, and the number of particles that can
be accelerated out of this population must be significantly smaller. The
scaling $\dot{N}\propto B_0\,n_0^{1/2}$ shows that this dilemma cannot be
overcome without supposing very high values of $n_0$ and $B_0$ (and $T_0$
--- to keep $t_R\sim1$~s) and, hence, increasing ${\cal P}_{\rm in}$ by a
huge amount. Even an increase of the average reconnection rate by one
order of magnitude could only weaken but not eliminate the problem.
Whereas a sufficient number of particles is available within the current
sheet to account for the decimetric radio source, it is likely that
acceleration in additional volumes contributes significantly to the flux
of HXR-emitting particles. Possibilities for this are (1) the spreading of
the acceleration volume by waves which accelerate particles stochastically
(e.g., Miller et al.\ \cite{Mi96}), or (2) the interaction between the
reconnection outflow and the low-lying loop, where the density is higher
(Tsuneta \cite{Ts95}), or (3) the transmission of magnetic stresses, which
are created in the current sheet by reconnection, along the field lines to
a more dense acceleration region near the base of the corona (Haerendel
\cite{Hae94}). The detailed correlation at small timescales between the
HXR and the decimetric emissions will then be relatively weak, at least
for possibilities (2) and (3), --- as is observed in the October 5, 1992
flare and in many other events (Aschwanden et al.\ \cite{Asch90}). 

Maximum energies of particles accelerated by the DC electric field near
X points have been
found to reach $W_{\rm max}\sim(m_e/m_i)^{1/4}e\delta_xE$ (in the ideal
case of test particle behaviour with no back-reaction on the
electromagnetic fields) (Kliem \cite{Kl94}; Kliem et al.\ \cite{KSS98}). 
This falls into the MeV range for all values of $n$, $T$, and $B$
considered here.  Thus, energies of radio-emitting particles
in flares ($\sim25\mbox{--}100$~keV) can be reached easily.

\subsection{Plasmoid evolution and frequency drift of the radio emission}
\label{drift}

The centre frequency of the decimetric continuum and pulses was drifting
on average toward lower frequencies during the whole event. This implies
in our model that the plasmoid was not stationary but moving upwards
during the whole radio burst so that the expansion at greater heights
could counteract the density increase due to reconnection
(Fig.~\ref{plasmoid}). If we extrapolate the observed height-time plot of
the plasmoid (Fig.~3 in Ohyama \& Shibata \cite{OS98}) backwards, based on
the assumption of constant acceleration from zero velocity at 
$t_0=\mbox{9:23:20~UT}$ to the observed velocity at 9:25~UT, we arrive at
a starting point, $h(t_0)\sim1.3\times10^4$~km, high above the top of the
flare loop [$h_{\rm fl}(t\!<\!\mbox{9:24:20~UT})\la3\times10^3$~km], which
is consistent with our simplified 2D model. 

Since $\beta\ll1$ in the corona, the  magnetic forces dominate the 
acceleration of plasmoids. In our model of plasmoid formation in
a preexisting current sheet, this implies that the reconnection rate at
the X point below the plasmoid exceeded the reconnection rate at the X
point above the plasmoid. The imbalance of the reconnected flux then led
to the upward bulk acceleration of the plasmoid. 

It has been found in previous non-symmetric simulations of dynamic current
sheet evolution (using a smaller box and partly different initial
conditions; Schumacher \& Kliem \cite{SK96}) that the reconnection rate is
most strongly enhanced during the buildup phase of a strong asymmetry,
while the bulk velocity of the plasmoid reaches its peak value somewhat
later (by $\sim50\,\tau_A$ in those simulations). It is thus expected that
the most prominent temporary increases of the reconnection rate are
associated with (1) enhanced particle acceleration, corresponding to
enhanced radio and X-ray flux, (2) an initial rise of pressure and density
in the plasmoid from the enhanced reconnection outflow into the plasmoid,
corresponding to a rise of the frequency of plasma emission, and (3) the
buildup of a strong asymmetry of the reconnected flux and, hence, a strong
acceleration of the plasmoid, corresponding to a strong negative drift of
the radio emission frequency in the inhomogeneous solar atmosphere,
following the initial rise of the emission frequency. The radio and hard
X-ray data suggest such an association during the intervals of strong flux
rise (near 9:24:30, 9:24:45, and 9:25:50~UT). The same effect is indicated
also at the beginning of the event (near 9:23:30~UT). 

The interplay between the increase of the density in the plasmoid by
reconnection and the decrease by upward motion and expansion can lead to a
constant or even decreasing frequency drift rate of the radio emission ---
in spite of increasing plasmoid velocity. Also the soft X-ray observations
suggest that the density of the plasmoid was only weakly decreasing during
its rise from $2.5\times10^4$~km at 9:25~UT to $5.2\times10^4$~km at
9:26~UT. Ohyama \& Shibata (\cite{OS98}) derived densities 
$n_{\rm pl}\sim(0.8\mbox{--}1.6)\times10^{10}$~cm$^{-3}$ and 
$n_{\rm pl}\sim(0.6\mbox{--}1)\times10^{10}$~cm$^{-3}$, respectively, at
these two instants (cf.\ their Tables 1 and 3). 

The very high frequency drift rates of the individual pulses were not
measured. An earlier statistical investigation of individual pulse drifts
at decimetric frequencies in 10 pulsation events has shown a narrow
distribution peaking at ${\rm d}f/{\rm d}t\approx3$~GHz\,s$^{-1}$,
consistent with compact sources of enhanced density and a scale height of
$\sim(2\mbox{--}20)\times10^3$~km  (Aschwanden \& Benz \cite{AB86}). We
note that the plasmoid observations during the October 5, 1992 flare as
well as our model are fully consistent with those source characteristics.
The weak statistical tendency of individual pulses toward positive
frequency drift is consistent with the density increase of the plasmoid
associated with each single coalescence event. 

Although the dynamical behaviour of the pulsating decimetric radio burst,
as well as the majority of the hard and soft X-ray data, appear to be
consistent with our model of the formation and acceleration of a (single)
large-scale plasmoid by dynamic magnetic reconnection, more complex
scenarios are not excluded. For example, it is possible that the pulsating
radio emission before 9:24:20~UT belonged to a separate plasmoid, which
may have formed independently if the current sheet had a large extent
along the line of sight ($z$ direction in the simulation). The drop of the
radio flux around 9:24:10~UT may perhaps be explained by such a multiple
source structure. Also the existence of two additional weak emission
bands, which showed signs of pulsations not or only weakly correlated with
the pulses of the main radio source (Fig.~\ref{threebands}), suggests the
formation of two additional radio sources (plasmoids separated from the
main plasmoid in the $x$ or $z$ direction, which are not resolved in the
soft X-ray images). 

The variations of the frequency drift rate may have been influenced also
by acceleration or deceleration of the plasmoid by the expanding loop, if
it was in fact connected with the plasmoid.

\section{Discussion}
\label{discuss}

The aspect ratio $r_{\eta}=\delta x_\eta/\delta y_\eta$ reaches $\sim10^2$
in the simulation.  The estimates of Sect.~\ref{scales} suggest that
$r_\eta\ga10^3$ in the corona.  This suggests that the modulations of the
reconnection rate, $R(t)$, become even deeper than in the simulation, since
$R$ is reduced more strongly during the phases of Sweet-Parker
reconnection.  This is in favour of deep modulations of the particle
acceleration rate like that observed in the flare under study.

Anomalous resistivity is created by kinetic current-driven instabilities,
which have the ion cyclotron radius, $r_{ci}$, and the ion inertial length,
$d_i$, as natural scales.  For our external plasma parameters $B_0$, $n_0$,
and $T_0$, for example, we have $r_{ci}\sim20$~cm and $d_i\sim700$~cm.
Thus, it is probable that the kinetic processes lead to small-scale
($L\ll\lambda_{\rm mfp}$) and short-lived ($\tau\ll t_R$) inhomogeneities
of the anomalous resistivity.  This may enforce intermittent excitation of
small-scale magnetic islands.  Such small-scale structures may be
destroyed again by diffusion or may merge into larger islands and enter
finally into the coalescence and plasmoid formation process on scales at
which a fluid description is appropriate.

More complicated schemes of dynamic reconnection are conceivable, based on
quasi-periodicities in complex multiple island coalescence processes (Kliem
\cite{Kl88}).  In that paper it was suggested that new islands are not
formed as single entities, but that excitation of the tearing mode in the
elongated flat sections of the current sheet creates a chain of islands.
The subsequent evolution may be rather similar to the dynamics obtained
here.  Since the initial phase of the coalescence instability is an ideal
MHD process, this instability develops faster than the tearing instability.
Therefore, in the presence of perturbations, it takes over as the nonlinear
evolution of the tearing instability (inhibiting saturation of the tearing
instability to a regular chain of islands).  The multiple coalescence of
the island chain in a section of the current sheet leads to formation and
amplification of plasmoids, similar to the evolution found here; it may
even be more dynamic.  Cyclic repetition of the sequence of tearing
instability, coalescence instability, and re-creation of long, flat current
sheet sections between plasmoids may result in pulsed particle
acceleration and radio emission.  Some aspects of such a complex
reconnection process have been studied by Schumacher \& Kliem
(\cite{SK96,SK97}).

Also externally triggered magnetic reconnection can show secondary tearing.
This was found by Odstr\v{c}il \& Karlick\'y (\cite{OK97}) for the case of
triggering by a weak shock wave incident on a current sheet.

Alternatively, or in addition to variable particle acceleration, the
pulses of the radio flux may be caused by varying efficiency of the
emission process in the disturbed and highly inhomogeneous plasma of the
dynamic current sheet.  For example, the irregular oscillations of the
central plasmoid, which were mentioned in Sect.~\ref{dynamic} and are
indicated by the variations of its peak pressure in Fig.~\ref{plasmoid},
could influence the radio emissivity.  However, varying particle
acceleration is a natural and immediate consequence of dynamic
reconnection.  A quantitative investigation of varying emission efficiency
is made difficult by insufficient knowledge of the distribution function
of the energetic particles and, hence, of the kinetic plasma physics of
the emission process.

An interpretation of the weak anticorrelation between the decimetric
pulses and the HXR lightcurve at short timescales (Fig.~\ref{cc}) in terms
of our current sheet model is possible but remains speculative.  A
tentative interpretation may be based on the oscillations of the central
plasmoid: The outflow of plasma and accelerated particles from the main X
points into the plasmoid is hampered, while the outflow away from the
plasmoid is supported, during phases of plasmoid expansion in the $x$
direction; and the situation is reversed during plasmoid expansion in the
$y$ direction. Alternatively, the anticorrelation could, for example, be
related to disturbances of the radio source caused by temporarily enhanced
particle acceleration in other parts of the active region (cf.\
Sect.~\ref{scales}). Studies of a larger sample of events will be required
to confirm the anticorrelation as a general property and to determine
possible causes. 

A model in which the pulsating radio burst is associated with a plasmoid
implies a strong role of trapping of downward propagating particles in the
cusp between the current sheet and the underlying flare loop. The weakness
of pulsations in the HXR burst is thus naturally explained. This would be
difficult if the radio source were formed within the flare loop, as is
commonly supposed for decimetric continuum bursts and was assumed for
the October 5, 1992 event by Aschwanden \& Benz (\cite{AB95}). 

An association of radio continuum burst sources with plasmoid ejections
had previously been found in case of isolated sources of moving type~IV
bursts at metric and decametric frequencies and much larger heights in the
corona ($\ga2\times10^{10}$~cm; e.g., Gopalswamy et al.\ \cite{Go97}). 
Recently, also Hori (\cite{Ho99}) suggested an association betwen drifting
decimetric/metric continuum bursts and the ejection of plasmoids for the
flares of October 5, 1992 and June 28, 1993. 

The proposed mechanism of variable particle acceleration may also underlie
quasi-periodic sequences of type III bursts (Aschwanden et al.\
\cite{Asch94}).  These events are also associated with the impulsive phase
of solar flares and their timescales are similar to those of the
pulsations considered here ($\sim0.5\mbox{--}5$~s).  A stronger role of
particle escape and a weaker influence of particle trapping may lead to
chains of type~III bursts instead of a pulsating radio continuum.  This
appears possible in case of strongly asymmetric reconnection, where the
field lines passing through the vicinity of the strongest X point below
the plasmoid do not immediately close at the X point above the plasmoid,
or in case of a sufficiently strong guide field component, $B_z>B_{x0}$,
where the particles can easily escape from the plasmoid into the $z$
direction. Since type~III burst sources generally follow open field lines,
the current sheet should extend into interplanetary space, i.e., exist
prior to the flare event in those cases.

\section{Conclusions}
\label{concl}

(1) We have presented a new model of pulsating radio bursts, in which the
variations of the flux are caused by modulations of the particle
acceleration in a highly dynamic reconnection process.  This process
operates in extended coronal current sheets and includes (in a fluid
model) the self-consistent evolution of anomalous resistivity.  It
involves the repeated formation of magnetic islands, their coalescence,
and the formation of one (or several) dense and hot plasmoid(s) which can
finally be ejected.  This model is able to explain the occurrence of
irregular or quasi-periodic pulses with timescales in the range
$\sim0.5\mbox{--}10$~s. It is also consistent with the usually low 
amplitude of modulation of the hard X-ray emission and with the observed 
weak anticorrelation between the hard X-ray and decimetric emissions at 
short timescales in the flare of October 5, 1992 considered here. 

(2) The unified explanation of plasmoid formation and pulsating radio
emission supports the conclusion by Ohyama \& Shibata (\cite{OS98}) that
the current sheet in the considered flare was not formed by the
ejection of the plasmoid (contrary to the often favoured view of eruptive
flare processes).  The model is consistent with both of the two remaining
possibilities --- formation and acceleration of the plasmoid within a
preexisting current sheet or simultaneous formation of both structures by a
global MHD instability.

(3) The emission mechanism of the radio waves may differ from the standard
model of radio continuum bursts in the decimetric and metric range, which
supposes particle injection and trapping in a stable coronal loop and
subsequent evolution of a loss-cone particle distribution.  Instead, a
large fraction of the accelerated particles may only temporarily be trapped
in the plasmoid, and the acceleration process itself may form an
anisotropic velocity distribution, which is unstable against electrostatic
or electromagnetic wave excitation.

(4) We have confirmed and extended previous numerical studies of impulsive
bursty reconnection, in particular the finding that magnetic reconnection,
which is triggered by anomalous resistivity, proceeds in a highly variable
manner if long current sheets are considered and the temporal and spatial
evolution of the anomalous resistivity is self-consistently taken into
account.

\begin{acknowledgements}
We gratefully acknowledge the very helpful comments by 
K.-L.~Klein and the anonymous referee. The Z\"urich radio observations are 
partially financed by SNF grant 20-53664.98. Technical support regarding 
the presentation of the Ond\v{r}ejov radio data was given by 
H.~M\'esz\'arosov\'a. This work was supported by DLR grant 50OC9706 and by 
the key projects
K1-003-601 and K1-043-601, and the grant A3003707 of the Academy of
Sciences of the Czech Republic. 
The John von Neumann-Institut f\"ur Computing, J\"ulich granted Cray
computer time. 
\end{acknowledgements}

\clearpage

\end{document}